\documentclass[ reprint,amsmath,amssymb,superscriptaddress,twocolumn]{revtex4-1}
\usepackage{graphicx}
\usepackage{dcolumn}
\usepackage{bm}
\usepackage{eucal}
\usepackage[version=3]{mhchem} 
\usepackage{url}
\usepackage{color}

\begin{document}
\title{Magnetic Bloch Oscillations and domain wall dynamics in a near-Ising ferromagnetic chain}

\author{Ursula B. Hansen*\email[]{uhansen@nbi.ku.dk}}
\affiliation{Niels Bohr Institute, University of Copenhagen, Universitetsparken 5, 2100 Copenhagen, Denmark}
\affiliation{Institut Laue-Langevin, CS 20156, 38042 Grenoble Cedex 9, France}

\author{Olav F. Sylju\aa sen}
\affiliation{Department of Physics, University of Oslo, P. O. Box 1048 Blindern, N-0316 Oslo, Norway}

\author{Jens Jensen}
\affiliation{Niels Bohr Institute, University of Copenhagen, Universitetsparken 5, 2100 Copenhagen, Denmark}

\author{Turi K. Sch\"affer}
\affiliation{Niels Bohr Institute, University of Copenhagen, Universitetsparken 5, 2100 Copenhagen, Denmark}

\author{Christopher R. Andersen}
\affiliation{National Centre for Nano Fabrication and Characterization, Technical University of Denmark, 2800 Kgs. Lyngby, Denmark}

\author{Jose A. Rodriguez-Rivera}
\affiliation{NIST Center for Neutron Research, National Institute of Standards and Technology, Gaithersburg, Maryland 20899, USA}
\affiliation{Department of Materials Science and Engineering, University of Maryland, College Park, MD 20742, USA}

\author{Niels B. Christensen}
\affiliation{Department of Physics, Technical University of Denmark, 2800 Kgs.\ Lyngby, Denmark}

\author{Kim Lefmann}
\affiliation{Niels Bohr Institute, University of Copenhagen, Universitetsparken 5, 2100 Copenhagen, Denmark}

\date{\today}
\begin{abstract} 
{\bf When charged particles in periodic lattices are subjected to a constant electric field, they respond by oscillating. Here we demonstrate that the magnetic analogue of these Bloch oscillations are realised in a one-dimensional ferromagnetic easy axis chain. In this case, the ``particle'' undergoing oscillatory motion in the presence of a magnetic field is a domain wall. Inelastic neutron scattering reveals three distinct components of the low energy spin-dynamics including a signature Bloch oscillation mode. Using parameter-free theoretical calculations, we are able to account for all features in the excitation spectrum, thus providing detailed insights into the complex dynamics in spin-anisotropic chains. } 
\end{abstract}
\maketitle
First described by F. Bloch in 1929~\cite{bloch_uber_1929}, electronic Bloch oscillations (BOs) are the oscillatory response of charged particles in a periodic potential to a constant electric field~\cite{Zener1934,Charleskittel2004}. The field gives rise to a force $\hbar\,dk/dt = qE,$ which drives the particle through the Brillouin zone. Upon crossing the Brillouin zone boundary the velocity $dE/dk$ is reversed leading to oscillatory motion. 
Observation of the resulting Bloch oscillations had to await the development of ultra pure semiconductor superlattices~\cite{mendez_stark_1988,Waschke1993,Leisching1994} and ultracold atoms in optical potentials~\cite{ben_dahan_bloch_1996,Wilkinson1996}. BOs have been observed directly in real space in waveguide arrays~\cite{Pertsch1999,Morandotti1999}, and more recently in a Bose-Einstein condensate~\cite{Geiger2018}. A magnetic analogue of the electronic BOs was predicted to exist in ferromagnetic near-Ising anisotropic spin-1/2 chains in a magnetic field~\cite{kyriakidis_bloch_1998}. In such chains, an excitation consisting of a domain wall (DW), separating regions of spins pointing up from spins pointing down, can be thought of as an analogue of charged particles in a periodic potential undergoing BOs in an electric field. The presence of anisotropic couplings is crucial for the magnetic Bloch oscillations (MBOs) in order to make a periodic band for the DW excitation. The magnetic field acts as a force trying to align spins, thus accelerating the DW in one direction, leading to an oscillatory motion in the same way as for the charged particle BOs. However, a single DW state is not stable in the bulk due to the large Zeeman energy cost of aligning many spins opposite to the field. 

Instead, states involving a pair of domain walls (2DW) bounding a short segment of overturned spins have been proposed as more favourable candidates for observation of magnetic Bloch oscillations~\cite{shinkevich_spectral_2012}. In this case the oscillation happens both for the DW at the beginning and for the DW at the end of a small cluster of adjacent spins aligned opposite to the magnetic field direction. 

The Hamiltonian of the ferromagnetic near-Ising anisotropic spin-1/2 chain in a longitudinal field is:
\begin{equation}
\mathcal{H} = \mathcal{H}^I + \mathcal{H}^a + \mathcal{H}^\perp, 
\end{equation}
where
\begin{subequations}\label{eq:Ham}
\begin{align}
	\mathcal{H}^I &= -\sum_i \mathcal{J}^z S^z_iS^z_{i+1} - g^z\mu_B\mu_0 H^z \sum_i S^z_i, \label{eq:HamI}\\
 	\mathcal{H}^a &= -\sum_i \mathcal{J}^a \left(S^+_iS^+_{i+1} +S^-_iS^-_{i+1} \right), \label{eq:Hama}\\
  	\mathcal{H}^\perp &= -\sum_i \mathcal{J}^\perp \left(S^+_iS^-_{i+1} +S^-_iS^+_{i+1} \right), \label{eq:Hamp}
\end{align}
\end{subequations}
\begin{figure*}
	\centering
		\includegraphics[width=0.8\linewidth]{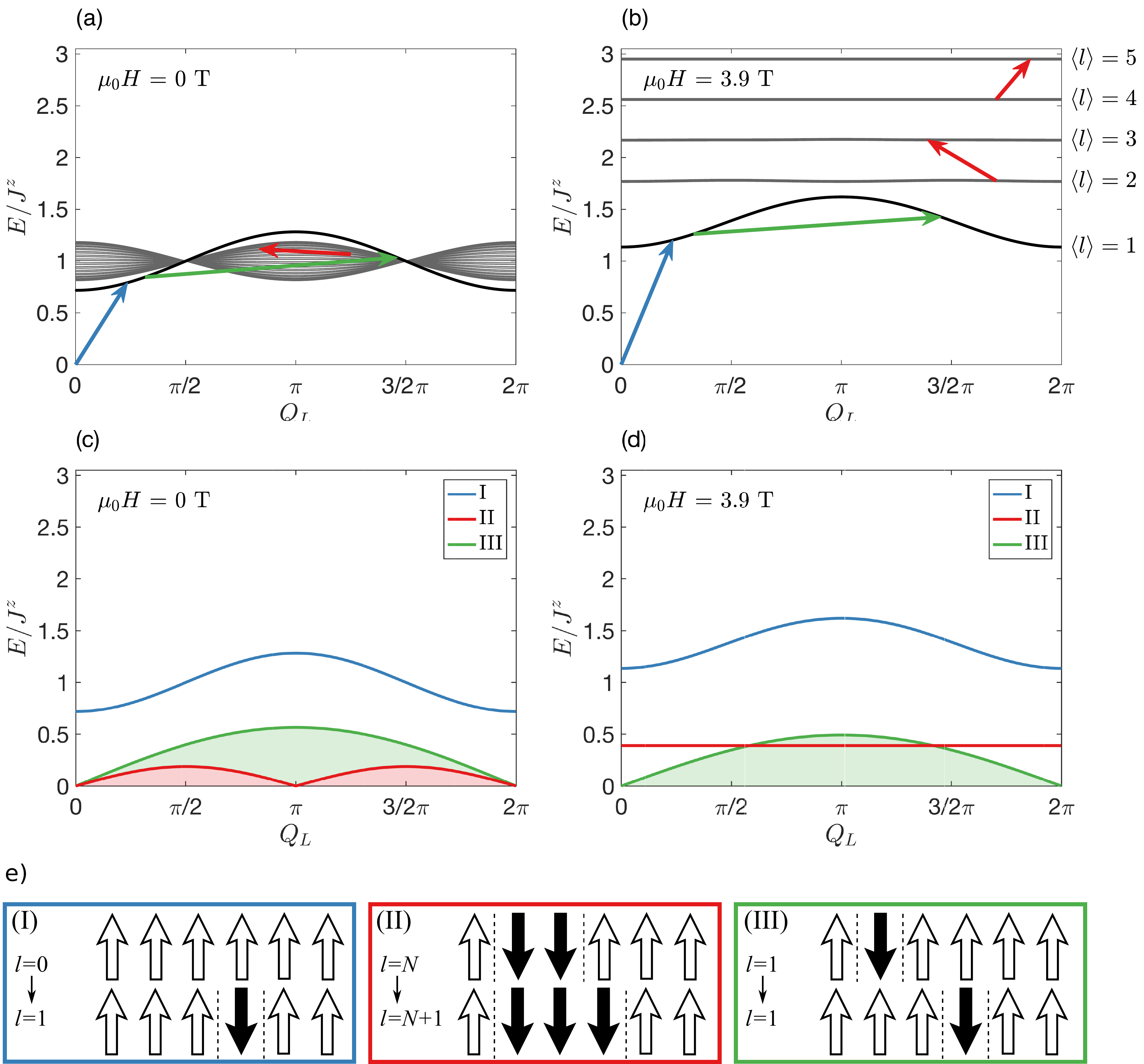}
	\caption[Cluster energy levels and magnetic excitations.]{{\bf Cluster energy levels and magnetic excitations.} Using the parameters $\mathcal{J}^a/\mathcal{J}^z = 0.05$, $\mathcal{J}^\perp/\mathcal{J}^z =  0.12$, $\mathcal{J}^z = 3.845$~meV and $g^z = 6.602$ relevant for a single chain in \ce{CoCl2*2D2O}, (a) and (b) show calculated cluster eigenenergies in zero field, $\mu_0H=0$, and in the high field limit where the WZL is well developed, $\mu_0H=3.9$~T. The energy of the ferromagnetic ground state is set to $E/\mathcal{J}^z=0$. (c) and (d) Interactions with neutrons cause three distinct types of transitions between the cluster eigenstates, shown by arrows in (a) and (b) and illustrated in (e), that each give rise to a contribution to the dynamic structure factor, shown in (c) and (d). (e)-(I): Creation of a $\langle l \rangle$=1 (single spin-flip) cluster leads to single-magnon scattering. (e)-(II): Transitions between cluster states characterized by different $\langle l \rangle$ can move the DW by one site. In zero field, this yields a continuum of periodicity $\pi$, while in a magnetic field it causes a well defined mode. (e)-(III) Intra cluster-level transitions for the $\langle l \rangle=1 $-cluster are reflected in a nearly field-independent continuum of period $2\pi$.}
	\label{fig:MBOsch}
\end{figure*}
with $\mathcal{J}^z>0$, $\mathcal{J}^a = (\mathcal{J}^x - \mathcal{J}^y)/4$ and $\mathcal{J}^\perp = (\mathcal{J}^x + \mathcal{J}^y)/4$. The usual spin raising and lowering operators are defined as $S_i^\pm = S_i^x \pm iS_i^y$.  The total effective $g$ factor along the $z$-axis is the sum of the orbital and spin $g$ factors: $g^z = g_L^z +g_S^z$. 

For an ideal Ising system ($\mathcal{J}^a = \mathcal{J}^\perp=0$) in zero magnetic field, the energy levels of all 2DW states, regardless the number of overturned spins, $l$, are degenerate, since the energy cost of creating a spin cluster comes only from the domain walls at either end of the cluster. In a magnetic field, the energy cost of aligning spins against the field splits the spectrum into a series of equidistant levels. In this case MBOs  do not occur: Since $[\mathcal{H},S^z_{\text{tot}}] = 0$, the cluster eigenstates all have fixed and time-independent numbers of spins oriented anti-parallel to the field. 

MBOs becomes possible when the cluster spectrum splits into equidistant levels in the presence of a magnetic field and anisotropic couplings ($\mathcal{J}^a \neq 0$), which cause the cluster  wave functions to become superpositions of cluster states of different lengths. Such a spectrum is known as the magnetic Wannier-Zeeman Ladder (WZL)~\cite{shinkevich_spectral_2012}, in full analogy with the electronic Wannier-Stark Ladder~\cite{Wannier1960}, which is the quantum mechanical signature of electronic BOs. When a WZL is formed, the expectation value of the cluster size $\langle l \rangle$ oscillates with a frequency $\omega_B$ corresponding to the energy difference between cluster states that both have contributions from spin clusters of the same size. In the present weak-anisotropy limit, this difference is given by the Zeeman energy:
\begin{equation}\label{eq:wb}
\hbar\omega_B = g^z\mu_B\mu_0 H^z.
\end{equation}

The existence of spin cluster excitations in \ce{CoCl2*2H2O} has already been demonstrated in far-infrared spectroscopy studies, affirming the predominant Ising nature of this system~\cite{Torrance1969,Tinkham1969}. The full energy spectrum of single cluster states (2DW) was calculated by one of us~\cite{shinkevich_spectral_2012} using parameters corresponding to a single chain in the deuterated, but magnetically identical material \ce{CoCl2*2D2O}. We  will here denote the class of states with mean integer cluster size $\langle l \rangle$ by $|\lambda_l\rangle$. In zero field the $|\lambda_{l=1}\rangle$  cluster state lies outside a continuum of $|\lambda_{l>1}\rangle$ cluster states with smaller bandwidths (Fig.~\ref{fig:MBOsch}(a)). In a sufficiently large magnetic field, the spin cluster continuum is split and the excitation spectrum now consists of the $|\lambda_{l=1}\rangle$  mode and the characteristic WZL (Fig.~\ref{fig:MBOsch}(b)), hence classifying \ce{CoCl2*2D2O} as a candidate material for observing MBOs. 

The intensity detected in a neutron scattering experiment has contributions from the correlation functions (see supplementary information)~\cite{JJ_rare}:
\begin{align}\label{eq:Sqw}
S^{\alpha\alpha}(Q,\omega ) = \frac{1}{Z}\sum_{{\lambda},{\lambda'}} \exp{(-E_{\lambda}/k_BT)} |\langle {\lambda'} | S_Q^\alpha |{\lambda}\rangle|^2\nonumber \\ \times \delta(\hbar\omega - (E_{\lambda'} - E_{\lambda})), 
\end{align}
where $S_Q^\alpha = \sum_{i} e^{i{Q}\cdot{R}_i } S^\alpha_i$ and $\alpha = x,y,$ or $z$. $Z$ is the partition function and $|{\lambda}\rangle$ and $|{\lambda'}\rangle$ are states of the system before and after the scattering event with energies $E_\lambda$ and $E_{\lambda'}$, respectively. Let us now consider how transitions between the cluster energy levels discussed above are reflected in three distinct contributions to the neutron scattering intensity. 
\begin{figure}[h!]
	\includegraphics[width=0.75\linewidth]{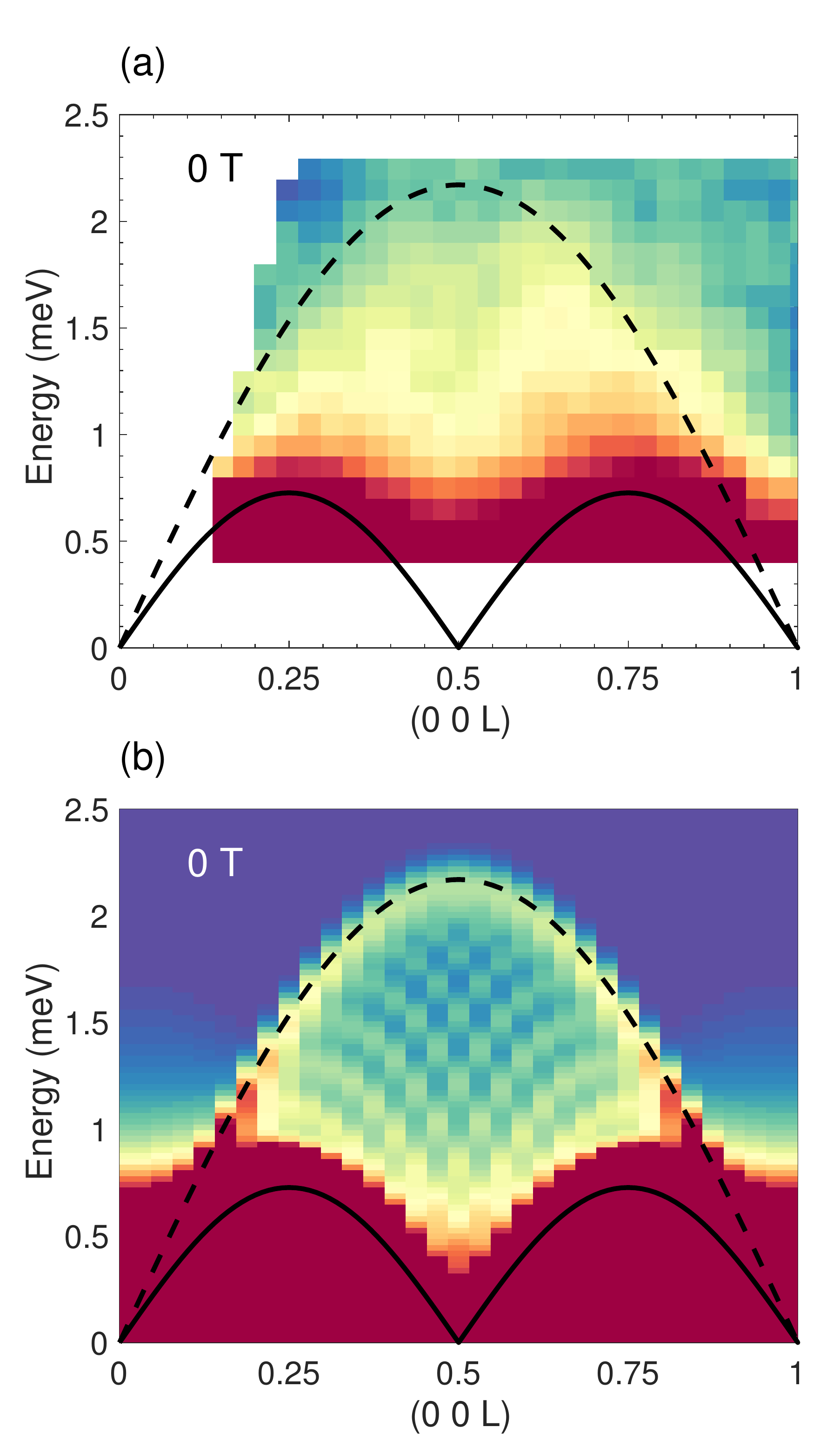}
	\caption{{\bf The zero field scattering continua.} (a) Inelastic neutron scattering intensity at zero field and $T=22$~K (b) Numerical approximation of $S(Q,\omega)$ using the same parameters as in Fig~\ref{fig:MBOsch}. Note that the chequerboard pattern within the continuum is due to finite size effects in the numerical calculation. Equation~\eqref{eq:dome_Ja} is superimposed as a black solid line and equation~\eqref{eq:dome_swsw} as a black dashed line.}
	\label{fig:Olav_num}
\end{figure}

At $T=0$,  only the ground state, $|\lambda_{l=0}\rangle$ with all spins parallel to $H$, is populated. The operators $S_Q^x$ and $S_Q^y$ cause transitions to states with a single spin flipped, as shown in Fig.~\ref{fig:MBOsch}(e)-(I). In turn, this state overlaps with the first excited 2DW state. The result is a dispersive spin wave excitation in $S(Q,\omega)$ which is found to appear in the range 4-6~meV in zero field~\cite{Kjems1975,montfrooij_spin_2001}. The spin wave contribution exhausts $S(Q,\omega)$ at $T=0$~K, since transition matrix elements between the ground state and higher lying cluster states (i.e. those dominated by $l>1$ contributions) are vanishingly small~\cite{shinkevich_spectral_2012}. In particular, it is not possible to probe the WZL, and hence MBOs, at low temperatures. However, when the temperature is increased, the states of the WZL will be thermally populated. The spectrum, is now richer and contains two additional contributions from transitions between \emph{different} excited 2DW states (Fig.~\ref{fig:MBOsch}(e)-(II)), and between different momentum states for the \emph{same} 2DW state (Fig.~\ref{fig:MBOsch}(e)-(III)). The first of these involves a change in the length of the spin cluster $|\lambda_{l}\rangle \rightarrow |\lambda_{l+1}\rangle$ caused by $S_Q^x$ and $S_Q^y$. For sufficiently large fields, this is reflected in $S(Q,\omega)$ in the appearance of a low energy peak at $\hbar\omega = \hbar\omega_B$~\cite{shinkevich_spectral_2012} signifying a WZL spectrum and the existence of MBOs. By contrast, in zero field, the process in Fig.~\ref{fig:MBOsch}(e)-(II) gives rise to a low energy transverse continuum. The upper boundary of this continuum is determined by the bandwidth, $4\mathcal{J}^a$, of the cluster states in Fig.~\ref{fig:MBOsch}(a), and can be approximated by:
\begin{equation}\label{eq:dome_Ja}
E_{\textup{II}}= 4\mathcal{J}^a|\sin(2\pi Q_\text{L})|, 
\end{equation} 
where $Q_\text{L}$ is the momentum transfer component along the reciprocal lattice vector $c^*$. This is the ferromagnetic equivalent of the Villain mode in antiferromagnetic Ising chains~\cite{Villain1975,Yoshizawa1981,Nagler1982}. The second finite temperature contribution to $S(Q,\omega)$ involves transitions \emph{within} the first excited 2DW eigenstate, $|\lambda_{l=1}\rangle \rightarrow |\lambda_{l=1}\rangle$ (Fig.~\ref{fig:MBOsch}(e)-(III)), and introduces a \emph{second} continuum in the same low-energy range occupied by the MBO peak in finite field, and by the transverse continuum in zero-field. The upper limit of this continuum, which is due to $S_Q^z$ and hence longitudinal, is given by:     
\begin{equation}\label{eq:dome_swsw}
E_{\textup{III}}= A |\sin(\pi Q_\text{L})|, 
\end{equation} 
where $A$  takes the value $4\mathcal{J}^\perp \left(1+\left(\mathcal{J}^a/\mathcal{J}^\perp \right)^2 \right)$ in zero magnetic field
and approaches $4\mathcal{J}^\perp$ in a strong magnetic field. (See Supplementary Information for explanations of equations~\eqref{eq:dome_Ja} and \eqref{eq:dome_swsw}). Figures~\ref{fig:MBOsch}(b) and (d) summarise all contributions to $S(Q,\omega)$ in zero field and in a finite magnetic field, respectively, using parameters relevant to a single chain of \ce{CoCl2*2D2O}.

We now turn to present our experimental results obtained on the high-flux neutron spectrometer MACS optimised for the study of weak and diffuse contributions to $S(Q,\omega)$~\cite{rodriguez_macsnew_2008}. The chosen setup (see Methods) improves on two previous unsuccessful attempts at identifying MBOs in \ce{CoCl2*2D2O}~\cite{christensen_magnetic_2000,montfrooij_spin_2001}. Figure~\ref{fig:Olav_num} shows our experimental data taken at 22~K, which is slightly above the ordering temperature $T_N=17.2$~K~\cite{Kjems1975}, in zero magnetic field, along with a theoretical calculation of $S(Q,\omega)$ for a single chain of \ce{CoCl2*2D2O}~\cite{shinkevich_spectral_2012}. The model is based on the thermal population of cluster states, and evaluation of the scattering matrix elements in equation~\eqref{eq:Sqw}. It can be seen to reproduce the salient features of the experimental data: There is a period $2\pi$ continuum peaked at $(0\,0\,\frac{1}{2})$ and bounded by equation~\eqref{eq:dome_swsw} with a realistic prefactor of $A=2.17$~meV, given by the coupling constants for a single chain in \ce{CoCl2*2D2O}. At lower energies we observe a period $\pi$ continuum peaked at $(0\,0\,\frac{1}{4})$ and $(0\,0\,\frac{3}{4})$, and bounded by a curve similar to equation~\eqref{eq:dome_Ja}, but with a prefactor slightly exceeding the expectation.   

Since the analytical calculations in Ref.~\onlinecite{shinkevich_spectral_2012} do not take into account the significant interchain couplings present in \ce{CoCl2*2D2O}, the quantitative details of the in-field inelastic neutron scattering data will henceforth be compared to numerical calculations based on a mean field/random phase approximation (RPA). This model considers a short chain segment of six neighbouring spin-$1/2$ and the spin Hamiltonian established in~Ref.~\onlinecite{JJ_RPA}. The model accounts for the bulk magnetic properties~\cite{Narath1965,Mollymoto1980} and describes both the spin waves~\cite{Kjems1975,montfrooij_spin_2001,christensen_magnetic_2000} and the magnetic cluster excitations~\cite{Torrance1969}. Further, it has successfully been applied to describe the quantum criticality in \ce{CoCl2*2D2O} in a transverse magnetic field~\cite{Larsen2017}. Within this model, interactions with neighbouring chains contribute to the effective field experienced by each domain wall. This leads to a modification of the Bloch energy,  equation~\eqref{eq:wb}, which for the case of \ce{CoCl2*2D2O} becomes:
\begin{equation}\label{eq:wB_eff}
\hbar\omega_B^* = g^z\mu_B \mu_0 H^z + \left(4\mathcal{J}^z_1+4\mathcal{J}'{}^{z}_1+2\mathcal{J}^z_2\right) \langle S^z \rangle.
\end{equation}
Here $\langle S^z \rangle$ is the average polarisation of the spin chains along the field axis. The values of the antiferromagnetic Ising-type interchain couplings $\mathcal{J}^z_1$, $\mathcal{J}'{}^{z}_1$ and $\mathcal{J}^z_2$ were determined in~Ref.~\onlinecite{JJ_RPA}  and are fixed throughout this paper (more details can be found in Supplementary Information).

\begin{figure}[h!]
	\includegraphics[width=0.75\linewidth]{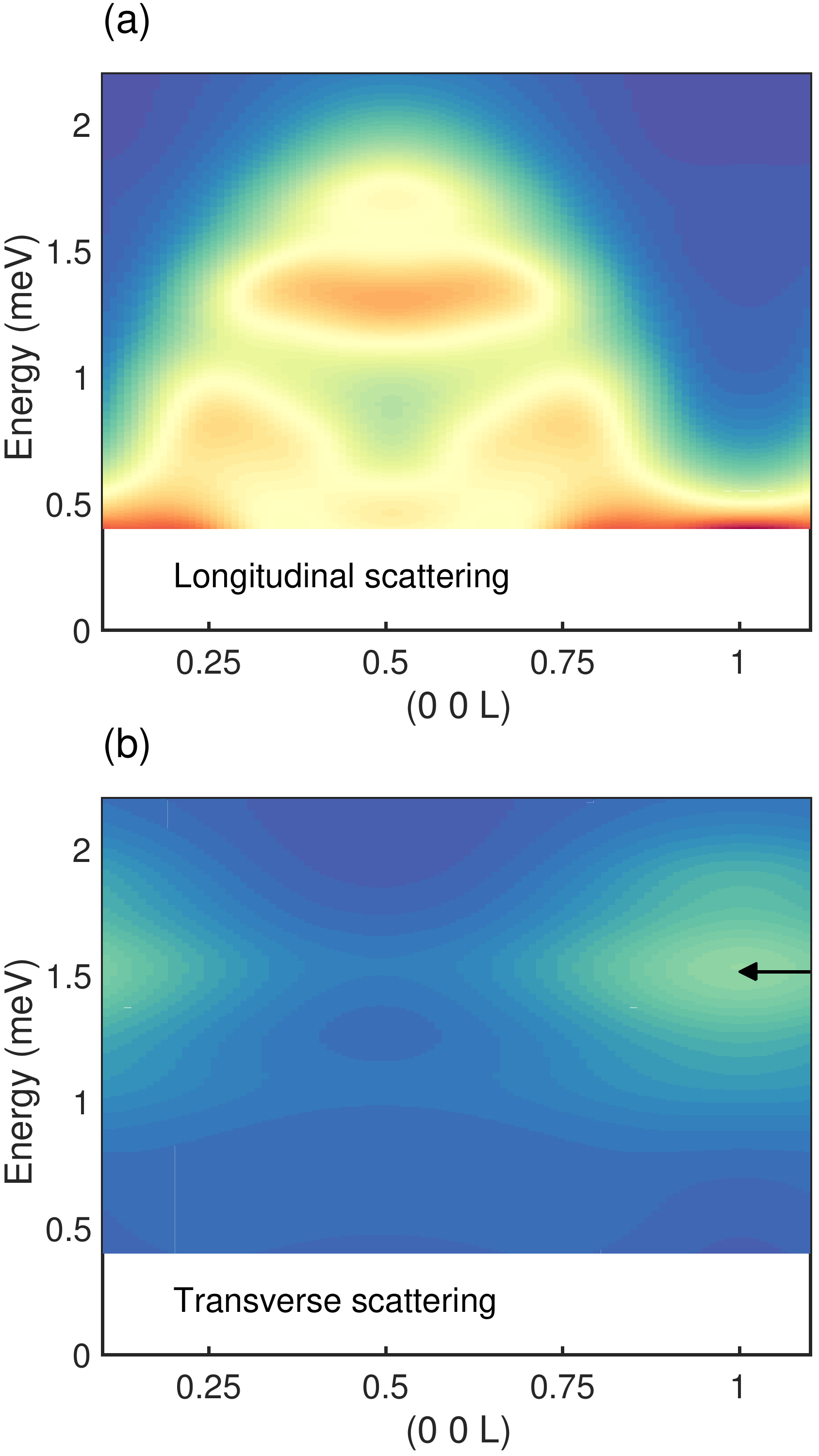}
	\caption{{\bf The transverse and longitudinal components of the excitation spectrum.} The calculated neutron scattering cross section in Fig.~\ref{fig:Sqw_RPA}d at 22~K and 7~T separated in its a) longitudinal  and b) transverse parts, proportional to $(g^z)^2 S^{zz}$ and $0.65 (g^x)^2 S^{xx} + 0.35 (g^y)^2 S^{yy}$, respectively (see supplementary information). The black arrows correspond to the prediction of the effective Bloch energy, $\hbar\omega_B^*$, given in equation~\eqref{eq:wB_eff}. The colour scale is the same for the two plots. }
	\label{fig:Sqw_LT}
\end{figure}
In keeping with our discussions based on Fig.~\ref{fig:Olav_num} and equation \eqref{eq:Sqw}, the RPA model predicts that the Bloch mode is transverse, while the continuum scattering is longitudinal. This is illustrated by Fig.~\ref{fig:Sqw_LT}. In addition the calculations highlight that although the MBOs corresponds to transitions between different levels of the WZL, which are approximately non-dispersive in the high field limit, the neutron scattering intensity nevertheless depends on the component of the neutron momentum transfer along $c^*$ with clear maxima at integer values of $Q_L$.

\begin{figure*}[t]
	\includegraphics[width=\linewidth]{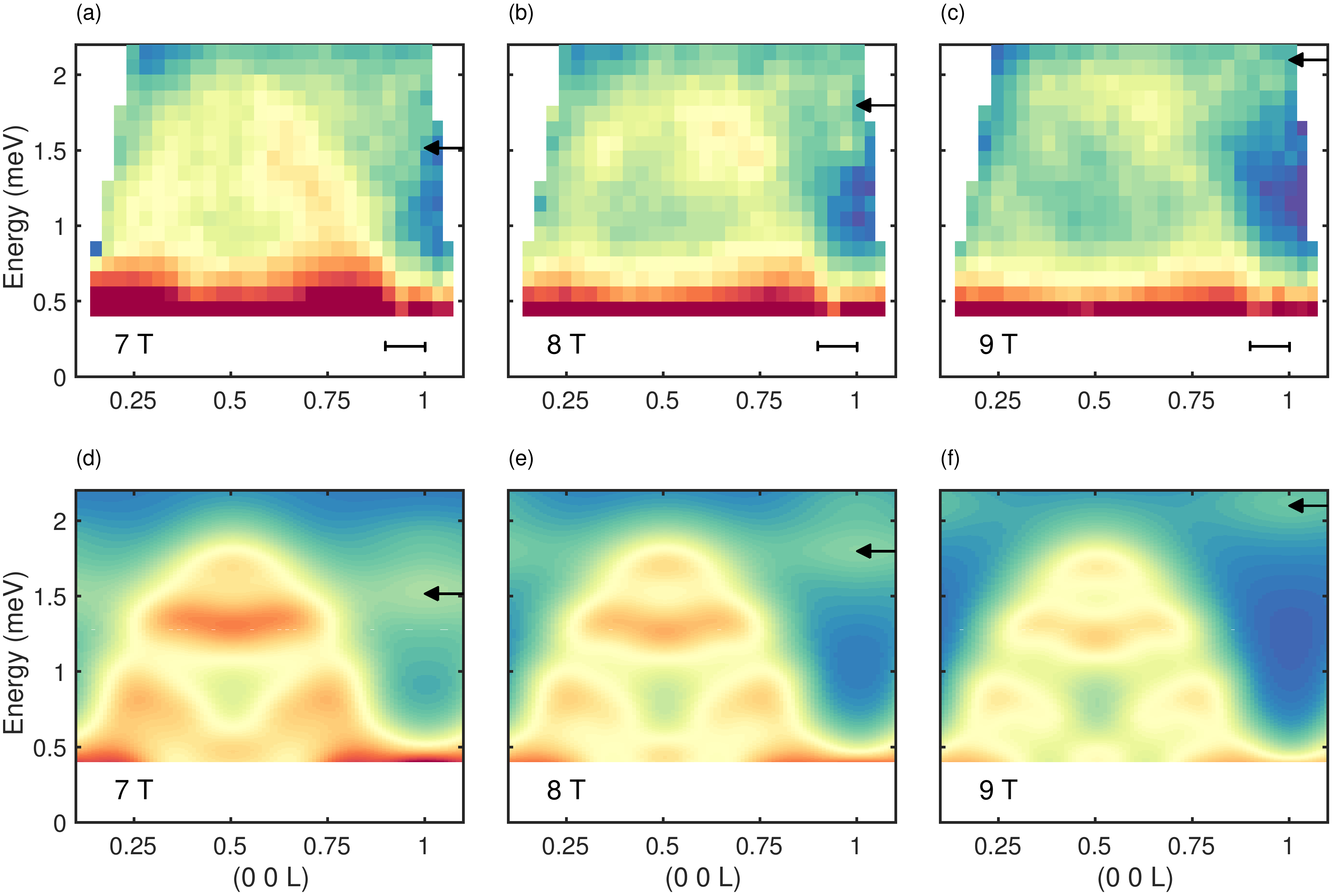}
	\caption{{\bf Inelastic neutron scattering data compared to RPA-calculations.} (a)-(c)  Experimental data collected at 22~K in magnetic fields 7~T, 8~T and 9~T, respectively. (d)-(f)~The calculated inelastic neutron scattering cross section, leaving out the variation due to the magnetic form factor (see supplementary information), as determined by the six spin-1/2 cluster model~\cite{JJ_RPA}. The black arrows corresponds to the prediction of the effective Bloch energy, $\hbar\omega_B^*$, given in equation~\eqref{eq:wB_eff}. The black bar in (a)-(c) represents the data integration range used to produce the data in Fig.~\ref{fig:Sw}.}
	\label{fig:Sqw_RPA}
\end{figure*}
Fig.~\ref{fig:Sqw_RPA}(a)-(c) shows neutron scattering intensity maps as a function of energy and momentum transfer measured at 22~K for magnetic fields of 7~T, 8~T and 9~T, respectively. For comparison, Fig.~\ref{fig:Sqw_RPA}(d)-(f) show the corresponding calculated excitation spectra. For all three magnetic fields the overall agreement between data and calculations is clear. The spectra are dominated by a largely field independent continuum that peaks at $Q = (0\,0\,\frac{1}{2})$ originating from the scattering of the thermally populated $|\lambda_{l=1}\rangle$ states, as described in equation~\eqref{eq:dome_swsw}. Outside the boundaries of this continuum, and close to $(0\,0\,1)$, it is possible to discern a weaker and clearly field-dependent contribution to $S(Q,\omega)$, consistent with the expectations for magnetic Bloch oscillations. The model calculations, Fig.~\ref{fig:Sqw_RPA}(d)-(f) predict a weak mode moving to higher energies with increasing field strength. The same phenomenon can be seen in the corresponding experimental data. It is worth noting that as the field increases, the period-$\pi$ continuum, equation~\eqref{eq:dome_Ja}, is seen to lose intensity on both data and model calculations. 
  
\begin{figure}
\centering
\includegraphics[width=\linewidth]{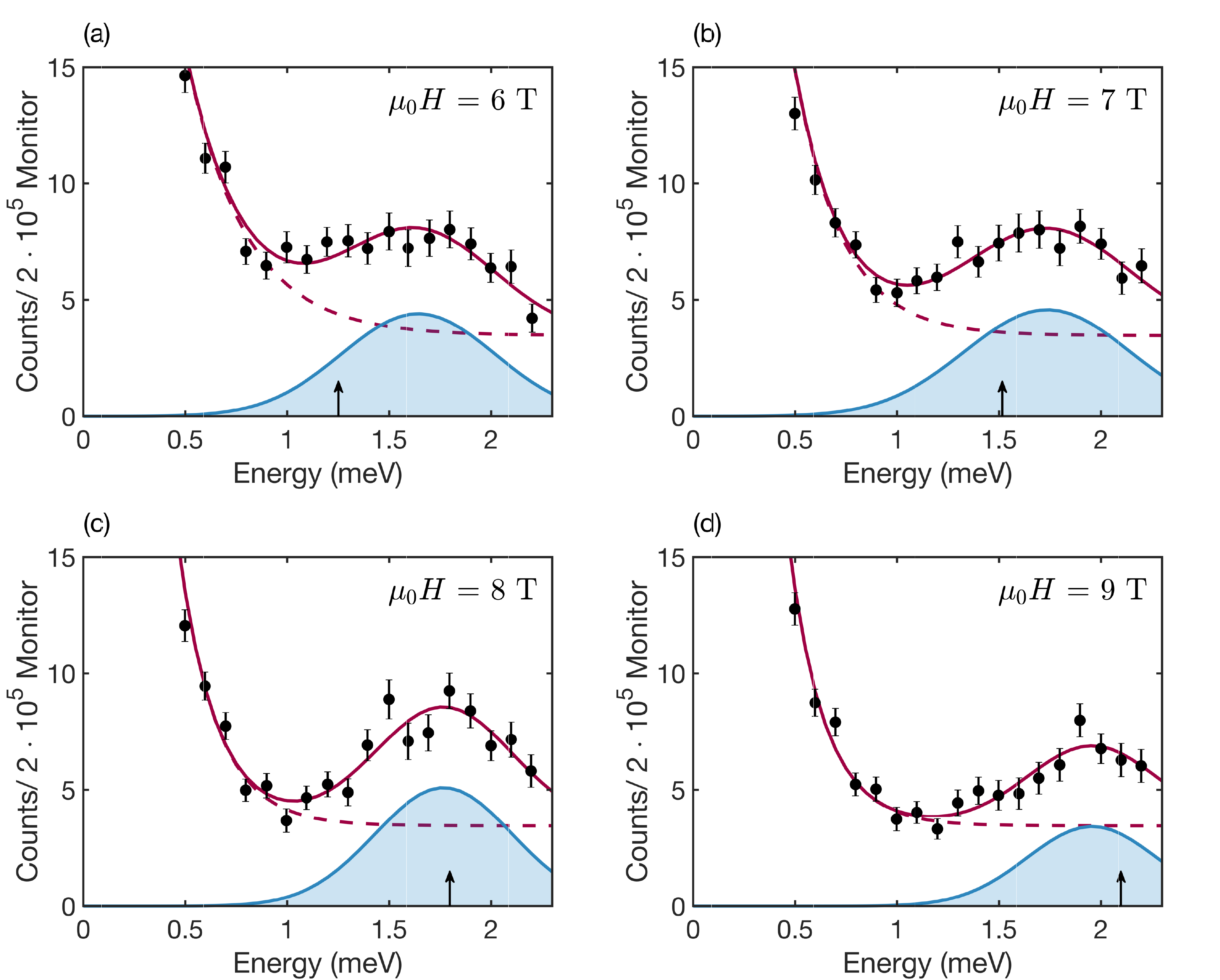}
\caption{{\bf Inelastic Neutron Scattering excitation spectra.} (a)-(d) Neutron scattering intensity averaged over the momentum range $Q_L=0.9-1.0$ r.l.u. for magnetic fields of 6~T, 7~T, 8~T and 9~T, respectively.  The red solid and red dashed lines correspond to the fit and background described in the main text and the shaded blue area to their difference - the MBO signal. Black arrows are the predicted effective Bloch energies, $\hbar\omega_B^*$, given in equation~\eqref{eq:wB_eff}.}
\label{fig:Sw}
\end{figure}

Fig.~\ref{fig:Sw} shows the evolution with magnetic field of the observed neutron intensity averaged over the momentum range $Q_L = 0.9 - 1.0$~r.l.u.. The signal that we attribute to the MBOs is clearly peaked and increases in energy with increasing field, while its amplitude approximately remains constant. (The accessible maximum energy transfer was constrained to 2.2~meV by the chosen experimental configuration, prohibiting us from tracking the MBO signal to higher fields). In order to isolate the signal originating from the MBOs, we subtract an effective background model reflecting incoherent scattering and a field-dependent contribution from the edge of the spin wave continuum. The resulting MBO signals are shown as blue areas in Fig.~\ref{fig:Sw}. 

\begin{figure}
\includegraphics[width=\linewidth]{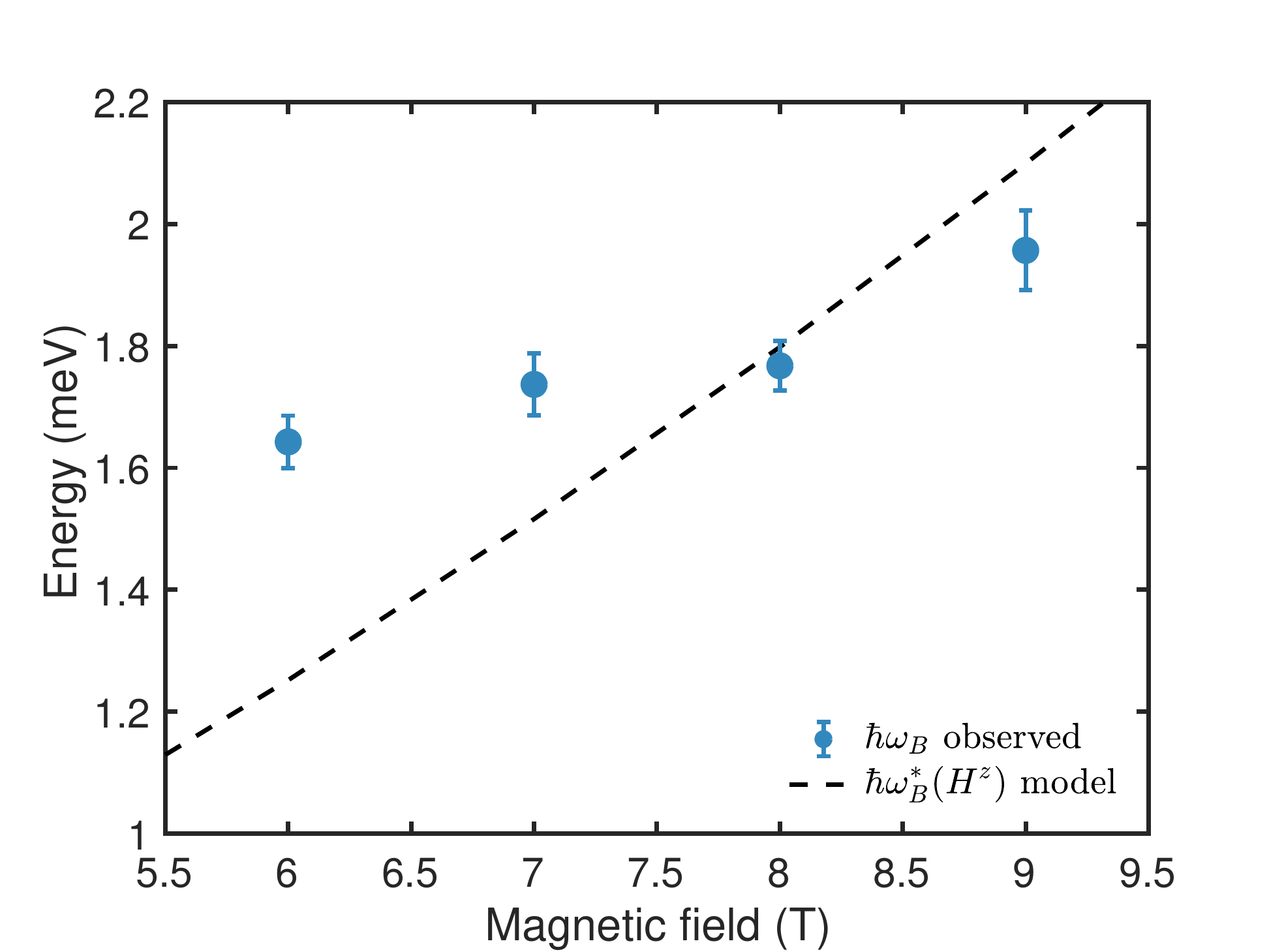}
\caption{{\bf Field dependence of the Bloch energy.} The position of the MBO peak for each of the energy scans shown in Fig.~\ref{fig:Sw}.  The error bars correspond to one standard deviation. The dashed line corresponds to the predicted Bloch energy $\hbar\omega_B^*$, equation~\eqref{eq:wB_eff}, which contains no free parameters.}
\label{fig:wB_H}
\end{figure}

In Fig.~\ref{fig:wB_H} we show the values of the Bloch energy obtained from the Gaussian fits in Fig.~\ref{fig:Sw} along with the predicted effective Bloch energy, equation~\eqref{eq:wB_eff}. The spin polarisation, $\langle S^z \rangle$, was calculated using the RPA model, which  has no free parameters. We see agreement for the highest fields, but a discrepancy between data and model for the two lowest fields. Decreasing the field leads to a lower polarization of the spin chains, which in turn means that the spins will be exposed to a broader distribution of effective fields due to a broader polarisation distribution of the neighbouring chains. In this case, a broader distribution of Bloch energies is to be expected. In addition, at lower magnetic fields, there will be a larger background contribution from the continuum which is difficult to separate from the signal coming from the MBOs. As a result, the systematic uncertainty on the Bloch frequency at 6 and 7~T, in reality will be larger than the standard deviation uncertainty from the fits shown as error bars in Fig.~\ref{fig:wB_H}. Resolving this discrepancy will require more detailed theoretical modelling and is beyond the scope of this work.   

 
In conclusion, we have studied the field-dependence of the low energy excitation spectrum in \ce{CoCl2*2D2O} at 22~K. We observe a broad field-induced peak close to the zone boundary, which is consistent with our model predictions of the signature of the MBOs. In spite of the discrepancies in the account of the field dependence of the MBO frequency, we consider the present experiments the first direct observation of Magnetic Bloch Oscillations. In addition to the Bloch Oscillations we have observed two thermally induced excitation continua that can be described using the anisotropic and perpendicular parts of the exchange couplings, $\mathcal{J}^a$ and $\mathcal{J}^\perp$.  The coherent reproduction of all three features in the experimental data, confirm our trust in the theoretical model and in our claim to have observed MBOs. Our results provide novel insights into the domain wall dynamics of anisotropic spin chains and add magnetic Bloch oscillations to the list of phenomena (including the spin Peierls transition~\cite{Boucher1996}, Dirac magnons~\cite{Pershoguba2018} and spinon Fermi surfaces~\cite{Shen2016}) initially introduced in the study of electrons in periodic solids, and subsequently observed to exist in model quantum magnets as well.

\section{Methods}
{\bf Sample preparation.} The single crystals of \ce{CoCl2*2D2O} were grown from a \ce{D2O}-solution by slow evaporation at 70~$^\circ$C. Their crystal structure were checked using x-ray diffraction and are consistent with literature values: $a= 7.256$~\AA, $b=8.575$~\AA, $c=3.554$~\AA\ and $\beta=97.6^\circ$, in the monoclinic space group C2/m $(\#12)$~\cite{Morosin1963}.  

{\bf Measurements.} The inelastic neutron scattering measurements were carried out at the MACS spectrometer at the NIST Center for Neutron Research~\cite{rodriguez_macsnew_2008}. Two single crystals of \ce{CoCl2*2D2O} with a total mass of 1.7~g  were co-aligned in the (H,0,L) plane on an aluminium holder. This sample was mounted in a vertical field cryomagnet allowing for magnetic fields up till 12~T along the (0,K,0)-direction. The final energy of the spectrometer was fixed at $E_f = 3.0$~meV. A Be filter was placed before the monochromator and a BeO filter after the sample. This configuration is optimal for studying weak signals but limits the neutron energy transfer to 2.2~meV. The scattered beam was analysed by a 20 channel detection system equipped with double-bounce PG analyser crystals~\cite{rodriguez_macsnew_2008}. The sample orientation with respect to the incident beam was fixed in all measurements. For each energy transfer, $S(Q,\omega)$ was probed for several orientations of the detection system. The background contribution from the empty cryomagnet was measured using the same configuration and subtracted from the signal obtained from the sample. 

{\bf Data availability.} All relevant data are available from the corresponding authors.
\section{Acknowledgements}
The project was funded by the Independent Research Fund Denmark through the project Spin Architecture and the Danish Agency for Science and Innovation through DANSCATT.  Access to MACS was provided by the Center for High Resolution Neutron Scattering, a partnership between the National Institute of Standards and Technology and the National Science Foundation under Agreement No. DMR-1508249. The project was further supported by the EU Interreg grant ``ESS \& MAX IV: Cross Border Science and Society" and Research Council of Norway, Grant No. 213606. We are thankful for assistance with crystal growth from H. Weihe.

\section{Author contributions}
U.B.H., C.R.A., J.A.R.-R., N.B.C. and K.L performed the inelastic neutron scattering experiment. U.B.H. analysed the data with help from K.L., N.B.C. and J.J. The analytical calculations were provided by O.F.S  and the RPA calculations by J.J. The samples were synthesized by T.K.S and aligned by U.B.H and C.R.A. The paper was written by U.B.H., with substantial contributions from O.F.S., J.J., N.B.C and K.L., and in discussions with all authors. 

\section{Competing interests} 
The authors declare no competing interests.


\begin{thebibliography}{01}
\bibitem{bloch_uber_1929} 
	Bloch, F.
	{\"Uber} die {Quantenmechanik} der {Elektronen} in {Kristallgittern},
	{\em Zeitschrift f\"ur Physik}, 
	52(7-8):555--600, 1929.

\bibitem{Zener1934}
Zener, C.
{A Theory of the Electrical Breakdown of Solid Dielectrics},
{\em Proceedings of the Royal Society A}, 
145(855):523--529, 1934.

\bibitem{Charleskittel2004}
	Kittel, C.
	{\em Introduction to Solid State Physics}.
    Wiley, 2004.
	
\bibitem{mendez_stark_1988}
	Mendez, E.~E., Agull\'o-Rueda, F. and Hong, J.~M. 
	{Stark} {Localization} in {GaAs}-{GaAlAs} {Superlattices} under an {Electric} {Field},
	{\em Physical Review Letters}, 60(23):2426--2429, 1988.
	
\bibitem{Waschke1993}
	Waschke, C., Roskos, H.~G., Schwedler, R.,  Leo, K., Kurz, H. and K{\"{o}}hler, K.
	{Coherent submillimeter-wave emission from Bloch oscillations in a	semiconductor superlattice}.
	{\em Physical Review Letters}, 70(21):3319--3322, 1993.
	
	\bibitem{Leisching1994}
	Leisching, P., {Haring Bolivar}, P., Beck, W., Dhaibi, Y., Br{\"{u}}ggemann, F.,
	Schwedler, R., Kurz, H., Leo, K. and K.~K{\"{o}}hler.
	{Bloch oscillations of excitonic wave packets in semiconductor superlattices}.
	{\em Physical Review B}, 50(19):14389--14404, 1994.
	
\bibitem{ben_dahan_bloch_1996}
	Ben~Dahan, M., Peik, E., Reichel, J., Castin, Y. and Salomon, C.
	Bloch {Oscillations} of {Atoms} in an {Optical} {Potential},
	{\em Physical Review Letters},  76(24):4508--4511, 1996.
	
\bibitem{Wilkinson1996}
	Wilkinson, S.~R., Bharucha, C.~F. , Madison, K.~W., Qian Niu, and Raizen, M.~G.
	{Observation of Atomic Wannier-Stark Ladders in an Accelerating Optical Potential}.
	{\em Physical Review Letters}, 76(24):4512--4515, 1996.
	
\bibitem{Pertsch1999}
	Pertsch, T., Dannberg, P., Elflein, W., Br{\"{a}}uer, A. and Lederer, F.
	{Optical Bloch Oscillations in Temperature Tuned Waveguide Arrays}.
	{\em Physical Review Letters}, 83(23):4752--4755, 1999.
	
\bibitem{Morandotti1999}
	Morandotti, R., Peschel, U., Aitchison, J.~S., Eisenberg, H.~S. and Silberberg, Y.
	{Experimental Observation of Linear and Nonlinear Optical Bloch Oscillations}.
    {\em Physical Review Letters}, 83(23):4756--4759, 1999.
	
\bibitem{Geiger2018}
	Geiger, Z. A., Fujiwara, K.~M., Singh, K. Senaratne. R., Rajagopal, S.~V., Lipatov, M., Shimasaki, T., Driben, R., Konotop, V.~V., Meier, T. and Weld, D. M.
 	{Observation and Uses of Position-Space Bloch Oscillations in an Ultracold Gas},
 	{\em Physical Review Letters}, 
	120(21):213201, 2018.
	
\bibitem{kyriakidis_bloch_1998}
	Kyriakidis, J. and Loss, D.
	Bloch oscillations of magnetic solitons in anisotropic spin-$1/2$ chains,
	{\em Physical Review B},  58(9):5568--5583, 1998.
	
\bibitem{shinkevich_spectral_2012}
	Shinkevich, S. and Sylju\aa sen, O.~F.
	Spectral signatures of magnetic Bloch oscillations in one-dimensional easy-axis ferromagnets,
	{\em Physical Review B}, 85:104408, 2012.
	
\bibitem{Wannier1960}
	Wannier, G.~H.
	Wave functions and effective hamiltonian for bloch electrons in an electric field,
	{\em Physical Review}, 117(2):432--439, 1960.
	
\bibitem{Tinkham1969}
	Tinkham, M.
	Microscopic Dynamics of Metamagnetic Transitions in an Approximately Ising System: \ce{CoCl2*2H2O},
	{\em Physical Review}, 188(2), 967-973, 1969.
	
\bibitem{Torrance1969}
	Torrance, J. B. and Tinkham, M.
	Excitation of Multiple-Magnon Bound States in \ce{CoCl2*2H2O},
	{\em Physical Review}, 187(2), 595-606, 1969.

\bibitem{JJ_rare}
	Jensen, J. and Mackintosh, A.~R.
	{Rare Earth Magnetism: Structures and Excitations}.
	{\em Clarendon Press, Oxford}, 1991.
	
\bibitem{Kjems1975}
	Kjems, K.~J., Als-Nielsen, J. and Fogedby., H.
	Spin-wave dispersion in \ce{CoCl2*2D2O}: A system of weakly coupled Ising chains,
	{\em Physical Review B}, 12(11):5190--5197, 1975.
	
\bibitem{montfrooij_spin_2001}
	Montfrooij, W., Granroth, G.~E., Mandrus, D.~G. and Nagler, S.~E.
	Spin dynamics of the quasi-one-dimensional ferromagnet \ce{CoCl2*2D2O},
	{\em Physical Review B}, 64(13):134426, 2001.
		
\bibitem{Villain1975}
	Villain, J.
	Propagative spin relaxation in the Ising-like antiferromagnetic linear chain,
	{\em Physica B+C}, 79(1):1--12, 1975.
	
\bibitem{Yoshizawa1981}
	Yoshizawa, H., Hirakawa, K., Satija, S.~K. and Shirane,  G.
 	Dynamical correlation functions in a one-dimensional Ising-like antiferromagnetic \ce{CsCoCl3}: A neutron scattering study.
	{\em Physical Review B}, 23(5):2298--2307, 1981.
	
\bibitem{Nagler1982}
	Nagler, S.~E., Buyers,  W.~J.L., Armstrong,  R.~L. and Briat,  B.
	Propagating domain walls in \ce{CsCoBr3}.
	{\em Physical Review Letters}, 49(8):590--592, 1982.
		
\bibitem{rodriguez_macsnew_2008}
	Rodriguez, J.~A., Adler, D.~M., Brand, P.~C., Broholm, C., Cook, J.~C., Brocker,  C., Hammond, R., Huang, Z., Hundertmark, P., Lynn, J.~W., Maliszewskyj, N.~C., Moyer, J., Orndorff, J., Pierce, D., Pike, T.~D., Scharfstein, G., Smee, S.~A. and Vilaseca, R.
	{MACS} - a new high intensity cold neutron spectrometer at {NIST},
	{\em Measurement Science and Technology}, 19(3):034023, 2008.
	
\bibitem{christensen_magnetic_2000}
	Christensen, N.~B., Lefmann, K., Johannsen, I. and J\o rgensen, O.
	Magnetic Bloch oscillations in the near-Ising antiferromagnet \ce{CoCl2*2D2O},
	{\em Physica B: Condensed Matter}, 276--278:784--785, 2000.
		
\bibitem{JJ_RPA}
	Jensen, J., Larsen, J. and Hansen, U.~B.
	{Comprehensive cluster-theory analysis of the magnetic structures and excitations in \ce{CoCl2*2H2O}}.
	{\em Physical Review B}, 97(2):024423, 2018.
	
\bibitem{Narath1965}
	Narath, A.
	{Antiferromagnetism in \ce{CoCl2*2H2O}. 2. Chlorine Nuclear Magnetic Resonance and Paramagnetic Susceptibility}.
	{\em Physical Review}, 140(2A):A552--A568, 1965.
	
\bibitem{Mollymoto1980}
	Mollymoto, H., Motokawa, M. and Date, M.
	{High Field Transverse Magnetization of Ising Antiferromagnet \ce{CoCl2*2H2O}},
	{\em Journal of the Physical Society of Japan}, 49(1):108--114, 1980.
		
\bibitem{Larsen2017}
	Larsen, J., Sch{\"{a}}ffer, T.~K., Hansen, U.~B., Holm, S.~L., Ahl, S.~R., Toft-Petersen, R., Taylor, J., Ehlers, G., Jensen, J., R{\o}nnow, H.~M., Lefmann, K., and Christensen, N.~B.
	{Spin excitations and quantum criticality in the quasi-one-dimensional Ising-like ferromagnet \ce{CoCl2*2D2O} in a transverse field}.
	{\em Physical Review B}, 96(17):174424, 2017.
	
\bibitem{Boucher1996}
Boucher, J.~P. and Regnault, L.~P.
\newblock {The Inorganic Spin-Peierls Compound \ce{CuGeO3}}.
\newblock {\em Journal de Physique I}, 6(12):1939--1966, 1996.

\bibitem{Pershoguba2018}
Pershoguba, S.~S., Banerjee, S., Lashley, J. C., Park, J., {\AA}gren, H.,
Aeppli, G. and Balatsky, A.~V.
\newblock {Dirac Magnons in Honeycomb Ferromagnets}.
\newblock {\em Physical Review X}, 8(1):011010, 2018.

\bibitem{Shen2016}
Shen, Y., Li, Y.-D., Wo, H., Li, Y., Shen, S., Pan, B., Wang, Q., Walker, H.~C.,
Steffens, P., Boehm, M., Hao, Y., Quintero-Castro, D.~L., Harriger, L.~W., 
Frontzek, M.~D., Hao, L., Meng, S., Zhang, Q., Chen, G. and Zhao, J.
\newblock {Evidence for a spinon Fermi surface in a triangular-lattice quantum-spin-liquid candidate}.
\newblock {\em Nature}, 540(7634):559--562, 2016.

\bibitem{Morosin1963}
Morosin, B. and Graeber, E.~J.
A reinvestigation of the crystal structure of \ce{CoCl2*2H2O},
{\em Acta Cryst}, 16, 1963.

\end{thebibliography}
\end{document}